\documentclass[twocolumn,showpacs,amsmath,amssymb]{revtex4-1}
\usepackage{graphicx}
\usepackage[squaren]{SIunits}
\usepackage{textcomp}
\usepackage{color}
\usepackage{subfigure}

\begin{document}

\title{Multicomponent exciton gas in cuprous oxide: cooling behaviour and the role of Auger decay}
\author{D.~Semkat}
\email{dirk.semkat@uni-rostock.de}
\author{S.~Sobkowiak}
\author{F. Sch\"one}
\author{H.~Stolz}
\affiliation{Institut f\"ur Physik, Universit\"at Rostock, 18051 Rostock, Germany}
\author{Th.~Koch}
\author{H.~Fehske}
\affiliation{Institut f\"ur Physik, Ernst--Moritz--Arndt--Universit\"at Greifswald, 17489 Greifswald, Germany}

\date{\today}

\begin{abstract}
In this paper we present a hydrodynamic model to describe the dynamics of para- and orthoexcitons in cuprous oxide at ultralow temperatures inside a stress induced potential trap. We take into account the finite lifetime of the excitons, the excitation process and exciton-phonon as well as exciton-exciton interaction. Furthermore, we model the two-body loss mechanism assuming an Auger-like effect and compare it to an alternative explanation which relies on the formation of biexcitons. We discuss in detail the influence on the numerical results and compare the predictions to experimental data.
\end{abstract}


\maketitle

\section{Introduction}

Bose--Einstein condensation (BEC) has been a field of growing interest in the last two decades, meanwhile proven in various systems including atoms, molecules, magnons, polaritons, etc. To establish BEC of excitons in a bulk semiconductor (proposed already in the 1960s \cite{moskalenko1962,blatt1962}), however, turned out to be a long standing problem that is still to be solved. Despite of many experiments even at extremely low temperatures (for an overview, see \cite{snoke2014}), so far conclusive evidence matching all required criteria \cite{snoke2003} could not be provided.

A promising candidate for the realisation of a BEC in a bulk material is cuprous oxide (Cu$_2$O) due to the long lifetime and the high binding energy of the excitons. In this material, some very promising signatures of an excitonic condensate have been found experimentally \cite{stolz2012}.

For the analysis of the reasons for the persisting discrepancy between the expectations and the experimental results, a critical revision of previous conceptions of the exciton gas physics is required.

In order to model the trapped exciton gas, the assumption of a global or local thermodynamic equilibrium, necessarily applied in previous analyses of the thermodynamics of the excitons \cite{stolz2010,sobkowiak2010,sobkowiak2011}, is not sufficient. During the measurements, the excitons undergo dynamic processes like spatial propagation towards the trap centre, conversion between ortho- and paraexcitons, and cooling by interaction with phonons. Therefore, a description of the kinetics of the excitonic system is required.

There exists a vast number of works on the kinetics of ultracold bosonic gases, for an overview see \cite{griffin2009}. Within the formalism of the well-known Zaremba--Nikuni--Griffin (ZNG) equations \cite{zaremba1999,griffin2009}, the dynamics of the condensed particles obey a generalised Gross--Pitaevskii equation containing the exchange between normal and condensed phases, while the evolution of the thermal particles is described by a quantum Boltzmann equation. To apply the formalism to the trapped exciton gas, semiconductor specific effects have to be taken into account by including collision terms between excitons and phonons as well as source and loss terms: laser excitation of excitons, their recombination, i.e., radiative decay, and a special two-body process usually referred to as Auger decay.
%The latter one will be discussed later on more in detail.

The first stage of the time evolution of the exciton gas is characterised by the local thermalisation of the excitonic momentum distribution. This process is very fast, within about 1 ns local equilibrium is reached \cite{sobkowiak2014}. The thermalisation turns out to be essentially determined by exciton-exciton collisions.

The cooling process of the excitons can be investigated looking at the ``effective'' temperature defined by the averaged kinetic energy per particle \cite{sobkowiak2014}. Again, the exciton-exciton collisions are crucial for the cooling efficiency. However, at very low temperatures of the helium bath below 0.1 K, the cooling is not efficient enough to ensure that the excitons reach the bath temperature within their lifetime \cite{sobkowiak2014}. This does not yet rule out the possibility that there are zones within the trap where the local temperature is less than the global effective temperature. The analysis of the local temperature distribution in the trap \cite{sobkowiak2015}, however, confirms the principle trend -- for temperatures above 0.1 K, the excitons reach bath temperature at least in the trap centre, while that is nowhere the case below that value. The temperature minimum is then outside the trap centre \cite{sobkowiak2015}. This behaviour is, on the one hand, caused by the drastically reduced cooling efficiency. On the other hand, the aforementioned Auger decay, a two-body process where one exciton recombines and the other one takes over the released energy, depends quadratically on the density and is thus a source of temperature increase in the trap centre where the exciton density has its highest value.

Thermodynamic as well as kinetic approaches to the exciton gas in the trap result in spatial (and temporal) profiles of exciton density and temperature. In order to compare with experimental findings, one has to translate the density distribution into the decay luminescence of excitons. Earlier investigations of the latter subject \cite{haug1983,shi1994} were based on an exciton-photon coupling Hamiltonian containing only ``normal'' terms (creating a photon while annihilating an exciton and vice versa). A recently published approach \cite{koch2016,koch2016a} based on the full minimal coupling Hamiltonian provides the excitation spectrum of the new quasiparticles composed of weakly interacting excitons (``bogolons'') and photons. However, it represents only a first step on the way to a general theory of excitonic decay luminescence.

The present analysis will focus mainly on the extension of the approach presented in \cite{sobkowiak2015} in two different aspects. First, the hydrodynamics will be reformulated for a multicomponent exciton gas. Second, the problem of the Auger-like two-body decay will be investigated more in detail, aiming both at a more elaborated calculation of the Auger coefficient as well as at comparing the impact of the Auger process on the hydrodynamic evolution to an alternative mechanism based on a transient biexciton formation \cite{wolfe2014}.

\section{Experimental Background}

The excitons under investigation are composed of holes in the $\Gamma_7^+$ valence band and electrons in the $\Gamma_6^+$ conduction band of Cu$_2$O (so-called yellow series). The ground state of this series consists of the non-degenerate paraexciton and the triply degenerate orthoexciton. The latter are labeled according to their spin projection ($+$), ($0$), and ($-$). The paraexcitons are the energetically lowest state lying $\unit{12.12}{\milli\electronvolt}$ below the orthoexcitons, due to electron-hole exchange interaction. Their long lifetime $\tau_\mathrm{P}=\unit{650}{\nano\second}$ \cite{schwartz2012} and high binding energy $E_\mathrm{B}^\mathrm{P}=\unit{151.36}{\milli\electronvolt}$ \cite{schwartz2012} make the paraexciton a promising candidate for a BEC of excitons in a bulk material.

Typical experiments in the field \cite{schwartz2012} are setup the following way. Excitons are created using a pump laser and collected inside a stress induced potential trap. The crystal specimen is cooled via a helium bath inside a cryostate. Helium bath temperatures as low as $\unit{37}{\milli\kelvin}$ have been reached using a $^3$He/$^4$He dilution cryostate \cite{stolz2012}, while optically pumping the crystal. The excitons themselves are cooled via interaction with the crystal lattice (phonons). The luminescence spectrum of the excitons is recorded using a CCD camera. 

There are different ways to create excitons experimentally. In a strain field one can excite ortho- or paraexcitons directly or create orthoexcitons indirectly under the involvement of a $\Gamma_3^{-}$-phonon. Due to its large oscillator strength the latter process is usually used \cite{schwartz2012,stolz2012}. The created orthoexcitons convert rapidly to paraexcitons at rates of $\unit{0.2}{\nano\second^{-1}}$ \cite{denev2002} to $\unit{0.29}{\nano\second^{-1}}$ \cite{wolfe2000}. The pump laser may be run in pulsed or continuous wave (cw) mode. Under pulsed excitation the system will eventually reach a quasi-equilibrium state which decays over time, while cw excitation results in a stationary state in which the creation and the decay of excitons balances out.

The stress applied on the crystal has different effects on the exciton species. The potential for ortho($+$)-, ortho($-$)-, and paraexcitons is attractive, while being strongly repulsive for ortho($0$)excitons. However, for all species it has cylindrical symmetry and can be calculated from experimental data using contact mechanics.

\section{Model} 

For a theoretical description we extend the ansatz presented in Ref.\ \cite{sobkowiak2015} to incorporate multiple components. The model introduced in Ref.\ \cite{sobkowiak2015} is based on the ZNG equations \cite{griffin2009}, which describe ultracold atomic gases in nonequilibrium. In order to derive a set of equations for a multicomponent system we use similar assumptions as in the ZNG formalism. Analogous to the one-component case the Bose-field operator for each component $\hat{\psi}_i(\mathbf{r},t)$ is split into $\hat{\psi}_i(\mathbf{r},t) = \Phi_i(\mathbf{r},t) + \tilde{\psi}_i(\mathbf{r},t)$ with condensate wavefunction $\Phi_i(\mathbf{r},t)$ and fluctuation operator $\tilde{\psi}_i(\mathbf{r},t)$. The exciton-exciton (X-X) scattering is assumed to be $s$-wave. Hence, the interaction strength $g_{ij}$ is given by
\begin{equation}
 \label{eq:gij}
 g_{ij}=2\pi\hbar^2\left( \frac{1}{m_i}+ \frac{1}{m_j} \right) a_{ij}^s\,,
\end{equation}
with the $s$-wave scattering length $a_{ij}^s$ and the exciton mass $m_i$. Furthermore, we neglect all nondiagonal densities $\tilde{n}_{ij}(\mathbf{r},t)=\langle \tilde{\psi}_i^\dagger(\mathbf{r},t) \tilde{\psi}_j(\mathbf{r},t)\rangle$ with $i \neq j$ and all anomalous densities $\tilde{m}_{ij}(\mathbf{r},t)=\langle \tilde{\psi}_i(\mathbf{r},t) \tilde{\psi}_j(\mathbf{r},t)\rangle$. The mean-field potentials are therefore given by 
\begin{eqnarray}
 \label{eq:MeanPotential}
 U_i(\mathbf{r},t) &=& V^i_\mathrm{ext}(\mathbf{r})+2g_{ii} [n^c_i(\mathbf{r},t) +\tilde{n}_i(\mathbf{r},t)] \nonumber \\ &&+ \sum_{j \neq i} g_{ij} [n^c_j(\mathbf{r},t) +\tilde{n}_j(\mathbf{r},t)]\,,
\end{eqnarray}
with the condensate density $n^c_i(\mathbf{r},t)=|\Phi_i(\mathbf{r},t)|^2$ and the density of the thermal excitons $\tilde{n}_{i}(\mathbf{r},t)=\langle \tilde{\psi}_i^\dagger(\mathbf{r},t) \tilde{\psi}_i(\mathbf{r},t)\rangle$. Under these assumptions, the dynamics of the condensates are governed by generalised Gross--Pitaevskii equations (GGPE) of the form
\begin{eqnarray}
 \label{eq:GGPE}
 \mathrm{i}\hbar\frac{\partial \Phi_i(\mathbf{r},t)}{\partial t} &=& \bigg[-\frac{\hbar^2\nabla^2}{2m_i}+ U_i(\mathbf{r},t) - g_\mathrm{ii}n^c_i(\mathbf{r},t) \nonumber \\ &&-\mathrm{i} R_i(\mathbf{r},t) \bigg] \Phi_i(\mathbf{r},t)\,,
\end{eqnarray}
with the coupling terms $R_i(\mathbf{r},t)$, which transfer excitons into and out of the condensate. 
Assuming the mean field potentials $U_i(\mathbf{r},t)$ to be only slowly varying, one can transform the equation of motion for $\tilde{\psi}_i(\mathbf{r},t)$ into a quantum Boltzmann equation 
\begin{eqnarray}
 \label{eq:BoltzmannEq}
 \frac{\partial f_\mathbf{k}^i(\mathbf{r},t)}{\partial t}+\frac{\hbar\mathbf{k}}{m_i}\cdot\nabla_{\mathbf{r}} f_\mathbf{k}^i(\mathbf{r},t) && \nonumber \\ - \frac{1}{\hbar} \nabla_{\mathbf{r}} U^i(\mathbf{r},t) \cdot \nabla_{\mathbf{k}}f_\mathbf{k}^i(\mathbf{r},t)  
 &=& \frac{\partial f_\mathbf{k}^i(\mathbf{r},t)}{\partial t}\bigg|_\text{coll.} 
\end{eqnarray}
for the Wigner distribution function $f_\mathbf{k}^i(\mathbf{r},t)$.

In the original ZNG equations only particle-particle scattering terms are included in Eq.\ (\ref{eq:BoltzmannEq}). Similar to Ref.\ \cite{sobkowiak2015} we extend the collision term to incorporate semiconductor specific effects, e.g. exciton-phonon (X-Ph) interaction. Then, the collision term reads  
\begin{eqnarray}
 \label{eq:collision}
 \frac{\partial f_\mathbf{k}^i(\mathbf{r},t)}{\partial t}\bigg|_\text{coll.} &=& \sum_j \left[C_\mathrm{X-X}^{ij} + C_\mathrm{X_c-X}^{ij} \right] \nonumber \\ &+& C_\mathrm{X-Ph}^i +  C_\mathrm{Conv}^i + C_\mathrm{C-D}^i \, ,
\end{eqnarray}
where $C_\mathrm{X-X}^{ij}$ stands for inter- ($i \neq j$) and intra-species ($i=j$) X-X scattering involving only thermal excitons. $C_\mathrm{X_c-X}^{ij}$ is the corresponding term if condensed and non-condensed excitons are involved. The interaction with phonons is contained in $C_\mathrm{Ph}^i$ and the conversion of different exciton species into each other in $C_\mathrm{Conv}^i$. All processes that can create (e.g. pump laser) or destroy (e.g. finite lifetime) excitons are grouped in $C_\mathrm{C-D}^i$. The experimentally observed effective two-body loss mechanism is also contained in this collision term.

In the ZNG equations, the energy dispersion for the non-condensed excitons is taken to be Hartree--Fock-like, %and given by
\begin{equation}
 \label{eq:HFDisp}
 \varepsilon_\mathbf{k}^i(\mathbf{r},t)=\frac{\hbar^2k^2}{2m_i}+ U_i(\mathbf{r},t) \, .
\end{equation}
This is a good approximation as long as no or only small condensates occur \cite{griffin2009}. The ZNG equations using a Bogoliubov quasiparticle spectrum can be found in \cite{imamovic2001}.

Due to the experimental background some additional simplifications are justified. A condensate of excitons is extremely unlikely to occur in any other species than the paraexcitons. Therefore, we only consider the GGPE for the paraexcitons. Since the ortho($0$)excitons are pushed out of the trap by the repulsive potential, we do not directly consider them in our model. The trap potentials for ortho($+$)- and ortho($-$)excitons are almost identical. The interaction with paraexcitons and phonons is also the same for both species. Therefore, we combine the ortho($+$)- and ortho($-$)excitons into one component, arriving at an effective two-component system of para- and orthoexcitons. Furthermore, we neglect their small mass differences and use $m=\unit{2.6}{m_0}$ for all excitons ($m_0$ -- free electron mass). In typical experiments the number of paraexcitons is much higher than the number of orthoexcitons. Therefore, we expect the influence of para-ortho X-X scattering on the paraexcitons to be very small, and neglect inter-species X-X scattering. Then, the coupling term $R(\mathbf{r},t)$ remains the same as in Ref.\ \cite{sobkowiak2015} and the GGPE is only modified via the mean-field potential compared to the one-component case.

\subsection{Hydrodynamic equations}

Like in the one-component case, we rewrite the quantum Boltzmann equations (\ref{eq:BoltzmannEq}) in terms of the first three moments. This leads to two sets of hydrodynamic equations governing the evolution of particle density, momentum density, and kinetic energy density of the thermal excitons of each species. These quantities are given by 
\begin{eqnarray}
\label{eq:nvepsDef}
\tilde{n}_i(\mathbf{r},t)&=&\int \frac{d\mathbf{k}}{(2\pi)^3} f^i_\mathbf{k}(\mathbf{r},t) \, , \nonumber \\
m\tilde{n}_i(\mathbf{r},t)\mathbf{v}_i(\mathbf{r},t)&=&\int \frac{d\mathbf{k}}{(2\pi)^3} \hbar\mathbf{k} f_\mathbf{k}^i(\mathbf{r},t) \, , \nonumber \\
E_i(\mathbf{r},t) &=& \int \frac{d\mathbf{k}}{(2\pi)^3} \frac{\hbar^2k^2}{2m} f_\mathbf{k}^i(\mathbf{r},t) \, ,
\end{eqnarray} 
with the (normal) velocity $\mathbf{v}_i(\mathbf{r},t)$ of the non-condensed excitons. In order to derive the hydrodynamic equations one has to multiply Eq.\ (\ref{eq:BoltzmannEq}) with ($\varphi_0=1$, $\varphi_1=\hbar\mathbf{k}$, $\varphi_2=\hbar^2k^2/2m$) and integrate over the whole $\mathbf{k}$-space. 
Each set of hydrodynamic equations needs to be closed, since the next higher moment appears in the equation for the energy, respectively. This can be achieved by assuming the form of a partial local equilibrium for the distribution function $f_\mathbf{k}^i(\mathbf{r},t)= \tilde{f}_\mathbf{k}^i(\mathbf{r},t)$ given by
\begin{eqnarray}
 \label{eq:LocalPartialEquil}
 \tilde{f}^i_\mathbf{k}(\mathbf{r},t) &=& [e^{[(\hbar \mathbf{k}-m\mathbf{v}_i)^2/2m +U_i-\tilde{\mu}_i(\mathbf{r},t)]/k_\mathrm{B}T_i(\mathbf{r},t)}-1]^{-1} \nonumber \\
 &=& [e^{[\hbar^2\tilde{k}^2/2m - \tilde{\mu}_\mathrm{eff}^i (\mathbf{r},t)]/k_\mathrm{B}T_i(\mathbf{r},t)}-1]^{-1} \, ,
\end{eqnarray}
with Boltzmann's constant $k_\mathrm{B}$ and the space- and time-dependent temperature and chemical potential $T_i(\mathbf{r},t)$ and $\tilde{\mu}_i(\mathbf{r},t)$. As already shown in Refs.\ \cite{sobkowiak2014,sobkowiak2015} the relaxation in $\mathbf{k}$-space for the experimentally relevant parameters is very rapid. Starting with a nonequilibrium distribution, the form (\ref{eq:LocalPartialEquil}) is typically reached in less than $\unit{1}{\nano\second}$. This is very fast compared to the lifetime of the paraexcitons of $\unit{650}{\nano\second}$. Therefore, the excitons can be described using a hydrodynamic model, if the relaxation into partial local equilibrium is split off and treated separately. This has to be done for all newly created excitons and will be explained in detail when the corresponding collision terms are discussed.

\section{Collision terms}\label{sec:collision}

\subsection{Pump laser $C_\mathrm{Laser}$}

The pump laser creates orthoexcitons either directly or under involvement of a $\Gamma^\mathrm{-}_3$-phonon. While both processes may occur simultaneously, we only consider the latter due to its much higher oscillator strength. The energy balance for this indirect process reads
\begin{equation}
\label{eq:LaserEnergy}
E_\mathrm{Laser}- \left[E_\mathrm{G}-E_\mathrm{B}^\mathrm{O} \right] = \varepsilon_\mathbf{k}^\mathrm{O}+ E_{\Gamma^-_3}\,,
\end{equation}
with the laser energy $E_\mathrm{Laser}=h c/\lambda$, the band gap $E_\mathrm{G}=\unit{2.17208}{\electronvolt}$ \cite{kazimierczuk2014}, the orthoexciton energy (\ref{eq:HFDisp}), their binding energy $E_\mathrm{B}^\mathrm{O}=\unit{139.24}{\milli\electronvolt}$ \cite{uihlein1981}, and the phonon energy $E_{\Gamma^-_3}=\unit{13.49}{\milli\electronvolt}$ \cite{hoeger2006}. The laser wavelength used in the experiments is $\lambda=\unit{605.9}{\nano\meter}$ \cite{stolz2012,schwartz2012}. For these parameters, Eq.\ (\ref{eq:LaserEnergy}) can only be fullfilled inside an attractive potential, thus, the pump laser only creates ortho($+$)- and ortho($-$)excitons, but not ortho($0$)excitons, since their potential is strongly repulsive.

To calculate the relaxation into partial local equilibrium we solve a homogeneous Boltzmann equation with the collision terms given by Eq.\ (\ref{eq:collision}). At $t=0$ there are no paraexcitons and the distribution function for the orthoexcitons is Gaussian. Its maximum is given by the $\mathbf{k}$-value corresponding to the kinetic energy of the orthoexcitons determined by Eq.\ (\ref{eq:LaserEnergy}). The width follows from the spectral width of the pump laser. The initial distribution function is normalised to the density of the newly created excitons. From these calculations we can determine the density of ortho- and paraexcitons entering the hydrodynamic equations and their respective energy. It also follows that both components have reached a partial local equilibrium after $\unit{1}{\nano\second}$, which is consistent with earlier results obtained by using simpler models \cite{sobkowiak2015}.

Spatially, the laser spot is placed $\unit{100}{\micro\meter}$ along the $z$-axis below the trap minimum. It has a Gaussian shape in $z$- and $\rho$-direction with a width of $\unit{3}{\micro\meter}$. The normalisation is chosen to resemble the exciton generation rate of a pump laser with power $P_L$, assuming that half of the emitted photons create excitons \cite{schwartz2012}. The moments of $C_\mathrm{Laser}$ are (symbolically) given by  
\begin{eqnarray}
 \Gamma^{(0,i)}_\mathrm{Laser}&=&n_\mathrm{Laser}^i(\mathbf{r},t) \, , \nonumber \\
 \mathbf{\Gamma}_\mathrm{Laser}^{(1,i)} &=& \mathbf{0} \, , \nonumber \\
 \Gamma^{(2,i)}_\mathrm{Laser}&=&E_\mathrm{Laser}^i(\mathbf{r},t) \, .  
\end{eqnarray} 
%

%The highly energetic excitons created by the laser relax into a partial local equilibrium within the first $\nano\second$.  

\subsection{Exciton-phonon collisions ($C_\mathrm{Ph}$)}

The main cooling mechanism in the system is the interaction of excitons with acoustic phonons. Orthoexcitons may interact with both transversal acoustic (TA), as well as longitudinal acoustic (LA) phonons. Paraexcitons in a stress-free crystal may only interact with LA phonons. However, due to the applied stress, interaction with TA$_1$-phonons becomes possible \cite{sobkowiak2014}. All these processes can be modeled using a deformation potential interaction. The moments of the corresponding collision term are \cite{haug1997,sobkowiak2015} 
\begin{eqnarray}
\label{eq:PhononMoments}
 \Gamma_\mathrm{X-APh}^{(n,i)} &=&- \frac{\pi D^2_i}{\varrho v_s}\int \frac{d \mathbf{k}d \mathbf{k}'}{(2\pi)^6} |\mathbf{k}'-\mathbf{k}| [\varphi_n(\mathbf{k})-\varphi_n(\mathbf{k}')] \nonumber \\ 
 &\times& \bigg[f^i_{\mathbf{k}}f_{\mathbf{k}'-\mathbf{k}}^\mathrm{Ph}(1+f^i_{\mathbf{k}'})   -(1+f^i_{\mathbf{k}}) (1+f_{\mathbf{k}'-\mathbf{k}}^\mathrm{Ph})f^i_{\mathbf{k}'} \bigg] \nonumber \\ 
 &\times& \delta \left(\varepsilon_\mathbf{k}^i -\varepsilon_{\mathbf{k}'}^i +\hbar\omega_{\mathbf{k}'-\mathbf{k}} \right) \, ,
\end{eqnarray}
with the speed of sound $v_s$, the crystal density $\varrho = \unit{6.11\times10^3}{\kilogram/\meter^3}$ \cite{haug1997}, and the deformation potential $D_i$. The values for the speeds of sound are $v_s^\mathrm{LA} =\unit{4.5\times10^3}{\meter/\second}$ and $v_s^\mathrm{TA}=\unit{1.3\times 10^3}{\meter/\second}$ \cite{trauernicht1986}. The deformation potential for the interaction of paraexcitons with LA-phonons is $D^\mathrm{LA}_\mathrm{P}=\unit{1.68}{\electronvolt}$ \cite{reimann89}. The respective value for the interaction with TA-phonons depends on the applied stress \cite{sobkowiak2014}. For our calculations, we use $D^\mathrm{TA}_\mathrm{P}= \unit{0.235} {\electronvolt}$, which corresponds to a trap depth of about $\unit{2}{\milli\electronvolt}$. The deformation potentials for the orthoexcitons are $D^\mathrm{TA}_\mathrm{O}= \unit{0.247} {\electronvolt}$ \cite{sobkowiak2014} and $D^\mathrm{LA}_\mathrm{O}= \unit{1.7} {\electronvolt}$ \cite{ohara1999a}.

For our model we assume the phonons to be in equilibrium at the lattice temperature $T_\mathrm{Ph}$ at all times. Hence, the phonon distribution function is given by $f_{\mathbf{k}'-\mathbf{k}}^\mathrm{Ph} =[\exp(\hbar\omega_{\mathbf{k}'-\mathbf{k}}/ k_BT_\mathrm{Ph})-1]^{-1}$ with the phonon energy  $\hbar\omega_{\mathbf{k}'-\mathbf{k}}=\hbar v_s |\mathbf{k}'-\mathbf{k}|$.

Additionally to acoustic phonons we also consider the interaction of excitons with optical phonons. Assuming a Fr\"ohlich-type interaction the moments are given by \cite{haug1997}
\begin{eqnarray}
\label{eq:OptPhononMoments}
 \Gamma_\mathrm{X-OPh}^{(n,i)} &=&- \frac{2 \pi}{\hbar}\int \frac{d \mathbf{k}d \mathbf{k}'}{(2\pi)^6} \Xi^2(\mathbf{k}'-\mathbf{k}) [\varphi_n(\mathbf{k})-\varphi_n(\mathbf{k}')] \nonumber \\ 
 &\times& \bigg[f^i_{\mathbf{k}}f_{\mathbf{k}'-\mathbf{k}}^\mathrm{Ph}(1+f^i_{\mathbf{k}'})   -(1+f^i_{\mathbf{k}}) (1+f_{\mathbf{k}'-\mathbf{k}}^\mathrm{Ph})f^i_{\mathbf{k}'} \bigg] \nonumber \\ 
 &\times& \delta \left(\varepsilon_\mathbf{k}^i -\varepsilon_{\mathbf{k}'}^i +E_\mathrm{Ph} \right)\,,
\end{eqnarray}
with the squared matrix element 
\begin{equation}
 \Xi^2(\mathbf{k}'-\mathbf{k})=\frac{2\pi E_\mathrm{Ph}e^2}{4\pi\varepsilon_0}\left(\frac{1}{\varepsilon_\infty}-\frac{1}{\varepsilon_{\omega=0}} \right)\frac{(q_e-q_h)^2}{|\mathbf{k}-\mathbf{k}'|^2} \, .
\end{equation}
The terms $q_e$ and $q_h$ enter the matrix element due to central-cell corrections and are given by
\begin{eqnarray}
 q_e &=& \left[ 1 + a_X \alpha_h |\mathbf{k}-\mathbf{k}'|/2  \right]^{-2}\,, \nonumber \\
 q_h &=& \left[ 1 + a_X \alpha_e |\mathbf{k}-\mathbf{k}'|/2  \right]^{-2}\,,
\end{eqnarray}
with the mass factors $\alpha_e=m_e/(m_e+m_h)$ and $\alpha_h=m_h/(m_e+m_h)$. For the electron mass we use $m_e=\unit{1.0}{m_0}$ and for the hole mass $m_h=\unit{0.7}{m_0}$.

There are two relevant optical phonons, the LO$_1$- and LO$_2$-phonon. Their energies are $E_\mathrm{Ph,1}=\unit{18.9}{\milli\electronvolt}$ and $E_\mathrm{Ph,2}=\unit{82.5}{\milli\electronvolt}$, respectively.
% -$\epsilon_\infty$ \newline \indent
% -$\epsilon_{\omega=0}$
% $a_X=\unit{0{,}7}{\nano\meter}$

\subsection{Conversion ($C_\mathrm{Conv}$)}

Ortho- and paraexcitons can convert into each other under involvement of a TA-phonon. This can be modeled using a deformation potential interaction \cite{wolfe2000}. The energetic splitting of the para- and orthoexcitons is an important quantity in the conversion process.

In a strain-free crystal the orthoexcitons lie $\unit{12.12}{\milli\electronvolt}$ above the paraexcitons. Under strain this splitting $\Delta$ depends on space and the applied stress. Given the potential traps for ortho- $V_\mathrm{ext}^\mathrm{O}(\mathbf{r})$ and paraexcitons $V_\mathrm{ext}^\mathrm{P}(\mathbf{r})$, the splitting $\Delta$ is simply $\Delta=V_\mathrm{ext}^\mathrm{O}(\mathbf{r})-V_\mathrm{ext}^\mathrm{P}(\mathbf{r})$. In the experiments under consideration it varies between $\unit{7}{\milli\electronvolt}$ and $\unit{9}{\milli\electronvolt}$. This corresponds to temperatures between $\unit{80}{\kelvin}$ and $\unit{100}{\kelvin}$. The effect of this splitting is twofold. It poses as a barrier for the conversion from para- to orthoexcitons, only allowing high-energy paraexcitons to convert to orthoexcitons. For the inverse process it leads to highly energetic paraexcitons created from the conversion. Since typical exciton temperatures in the experiments are below $\unit{2}{\kelvin}$, we neglect the para-ortho conversion process. The $n$-th moments of the collision terms for the ortho-para conversion are given by \cite{wolfe2000}
\begin{align}
 \label{eq:OPnteMomentAllg}
 &\Gamma_\mathrm{Conv}^{(n,\mathrm{P})}=\frac{\pi L^2}{\rho v_\mathrm{TA}} \int \frac{\mathrm{d}^3k}{(2\pi)^3} \int \frac{\mathrm{d}^3k'}{(2\pi)^3} |\mathbf{k}-\mathbf{k}'| \Phi_n(\mathbf{k}) \\
 &\times f_{\mathbf{k}'}^\mathrm{O}(1+f_{\mathbf{k}}^\mathrm{P})\Big\{(1+f_{|\mathbf{k}-\mathbf{k}'|}^\mathrm{Ph})\delta(\varepsilon_{\mathbf{k}'}^\mathrm{O}-\varepsilon^\mathrm{Ph}_{|\mathbf{k}-\mathbf{k}'|} -\varepsilon_{\mathbf{k}}^\mathrm{P}) \nonumber \\ 
 & + f_{|\mathbf{k}-\mathbf{k}'|}^\mathrm{Ph} \delta(\varepsilon_{\mathbf{k}'}^\mathrm{O} +\varepsilon^\mathrm{Ph}_{|\mathbf{k}-\mathbf{k}'|} -\varepsilon_{\mathbf{k}}^\mathrm{P}) \Big\} \nonumber \\ 
 &\qquad\;\;= \frac{\pi L^2}{\rho v_\mathrm{TA}} \int \frac{\mathrm{d}^3k}{(2\pi)^3} \int \frac{\mathrm{d}^3k'}{(2\pi)^3} \Phi_n(\mathbf{k}) F(\mathbf{k},\mathbf{k}')\,, \nonumber \\
 &\Gamma_\mathrm{Conv}^{(n,\mathrm{O})}=-\frac{\pi L^2}{\rho v_\mathrm{TA}} \int \frac{\mathrm{d}^3k}{(2\pi)^3} \int \frac{\mathrm{d}^3k'}{(2\pi)^3} \Phi_n(\mathbf{k}') F(\mathbf{k},\mathbf{k}')\,, \nonumber
\end{align}
with deformation potential $L=\unit{50}{\milli\electronvolt}$ \cite{wolfe2000}.

\subsection{Lifetime ($C_\tau$)}

Both species of excitons have a finite lifetime. We model this via the collision term $C_\tau^i=-\tilde{f}_\mathbf{k}^i(\mathbf{r},t)/\tau_i$ with a constant lifetime $\tau_i$. The corresponding moments are
\begin{eqnarray}
\label{eq:Gammatau}
\Gamma_\mathrm{\tau}^{(0,i)}&=&-\tilde{n}_i(\mathbf{r},t)/\tau_i\,, \nonumber \\
\mathbf{\Gamma}_\mathrm{\tau}^{(1,i)}&=&-m\tilde{n}_i(\mathbf{r},t)\mathbf{v}_i(\mathbf{r},t)/\tau_i\,, \nonumber \\
\Gamma_\mathrm{\tau}^{(2,i)}&=&-E_i(\mathbf{r},t)/\tau_i \, .
\end{eqnarray} 
For the calculations we assume a paraexciton lifetime of $\tau_\mathrm{P}=\unit{650}{\nano\second}$ \cite{stolz2012} and an orthoexciton lifetime of $\tau_\mathrm{O}=\unit{150}{\nano\second}$ \cite{schwartz2012}. However, the effective lifetime of the orthoexcitons is limited rather by the rapid conversion to paraexciton, than by this process. 

\subsection{Two-body decay}
In experiments an effective loss mechanism, that scales with the square of the density, has been observed. Two different explanations have been put forward to explain this effect. One possibility is an Auger-like two-body decay of the excitons. In this process one exciton recombines, while ionising another. The second explanation attributes the loss mechanism to the formation of biexcitons, which in turn undergo an Auger-like decay themselves. Hence, in both cases two excitons are destroyed and an electron-hole pair is created. The latter rebinds to form a high-energy exciton. 

To model the Auger-like two-body decay we use a $\mathbf{k}$-averaged Auger rate $A$ as commonly employed to explain experimental data \cite{ohara1999,yoshioka2010,schwartz2012}. The loss of excitons of the $i$-th species can then be described by the collision term 
\begin{equation}
 C_\mathrm{Auger}^{i}=-2A_{ii}\tilde{n}_i(\mathbf{r},t)\tilde{f}^i_\mathbf{k}(\mathbf{r},t)- A_{ij}\tilde{n}_j(\mathbf{r},t)\tilde{f}^i_\mathbf{k}(\mathbf{r},t)\,,
\end{equation}
with constant Auger rates $A_{ij}$.

\subsubsection{Estimation of the Auger coefficient}
The Auger coefficient for the nonstressed crystal can be calculated based on \cite{baym1996}. Two possible decay channels are of particular interest: direct and phonon assisted Auger scattering. In an unstrained crystal both processes require the recombining exciton to be an orthoexciton. While symmetry allows the transition to orthoexcitons via quadrupole coupling, direct transition of paraexcitons is strictly forbidden. In case of the phonon-assisted transition, the creation of orthoexcitons under the involvement of a $\Gamma_3^-$-phonon is the dominant transition channel, while paraexcitons solely couple to the weak $\Gamma_5^-$-phonon mode, thus diminishing the recombination rate of paraexcitons greatly. Indeed, the oscillator strength of the $\Gamma_3^-$-phonon assisted transition is predominant, preferring its treatment.
The phonon-assisted scattering matrix can be expressed in second order perturbation theory, effectively splitting it into two separate processes: the phonon transition of a later recombining orthoexciton into an intermediate state and the subsequent direct scattering of the intermediate state exciton with a second exciton of indifferent species
\begin{eqnarray}\label{eq:fs01}
&&M_\lambda^\mathrm{(pa)} =\\
&&\frac{\langle \Phi_\mathrm{1Sy}(\mathbf{K})|h_{\nu,\mathbf{Q}} |\Phi_\lambda (\mathbf{K-Q})\rangle \;M_\lambda^\mathrm{(d)}(\mathbf{K-Q},\mathbf{P},\mathbf{k_e},\mathbf{k_h})}{E_{\mathrm{1Sy}}(\mathbf{K}) - E_\lambda (\mathbf{K}-\mathbf{Q})} \,.\nonumber
\end{eqnarray}
In this case, the phonon mode $\nu$ belongs to the $\Gamma_3^-$ phonon, and the intermediate state $\lambda$ is theoretically any S-exciton state of the blue series of Cu$_2$O. In \cite{schoene2017} it was shown that, for the intermediate blue series, states with principal quantum numbers beyond the 1S state do effectively not participate, thus the treatment of the blue 1S state suffices. The phonon transition element for nonpolar optical phonons is generally expressed via an optical deformation potential
\begin{eqnarray}\label{eq:ds02}
 &&\langle \Phi_\mathrm{1Sy}(\mathbf{K})|h_{3,\mathbf{Q}} |\Phi_\mathrm{1Sb} (\mathbf{K-Q})\rangle \nonumber \\
 &&\simeq \mathcal{S}^\mathrm{(1Sb,1Sy)}(Q)\, D_{3;68} (Q) \sqrt{\frac{\hbar}{2\rho\,\Omega\, \omega_{3}}}\,.
\end{eqnarray}
The first term $\mathcal{S}^\mathrm{(1Sb,1Sy)}(Q)$ is a convolution of the yellow and blue exciton 1S wave functions in momentum space. The second term $D_{3;68} (Q)$ is the optical deformation potential between the $\Gamma_6^+$ and $\Gamma_8^-$ conduction bands via the $\Gamma_3^-$ phonon. The square root contains the material density $\rho$, crystal volume $\Omega$ and the phonon energy $\hbar\omega_{3}$. For small phonon momenta $\mathbf{Q}$ the deformation potential can be expanded into a Taylor series as
\begin{equation}\label{eq:fs03}
 D_{\lambda,ij}(Q) = \; D_{\lambda,ij}^{(0)} \;+\; D_{\lambda,ij}^{(2)}\,Q^2\; + \ldots \;.
\end{equation}
This approach differs fundamentally from the treatment of \cite{baym1996} since intermediate states are assumed to be exciton states and the deformation potential is momentum dependent.
Similarly, the direct scattering matrix element $M_\mathrm{1Sb}^\mathrm{(d)}$ features the wave function of the intermediate 1S blue state. The direct scattering can be separated into two processes, one where the energy of the recombined exciton is transferred via Coulomb interaction to the electron, in the other case the energy is transferred to the hole. Both terms are additive and yield the collective direct scattering matrix element
\begin{eqnarray} \label{eq:fs04}
M^\mathrm{(d)}_\mathrm{1Sb} 
%= M^\mathrm{(d)e}_\mathrm{b} + M^\mathrm{(d)h}_\mathrm{b} \\
&=& \frac{\hbar}{m_0}\,V_\mathrm{eff}(\mathbf{K\!-\!Q})\\
&&\times\sum_\mathbf{q} \varphi^{1\mathrm{Sb}}_{\mathbf{q}-(\mathbf{K}-\mathbf{Q})/2} 
\frac{\langle u_{\mathrm{7v},\mathbf{q}}|(\mathbf{K}-\mathbf{Q})\cdot \mathbf{p}| u_{\mathrm{8c},\mathbf{q}}\rangle}{E_\mathrm{8c}(\mathbf{q}) - E_\mathrm{7v}(\mathbf{q})} \nonumber \\
&&\times  \left[ \varphi^{\mathrm{1Sy}}_{\mathbf{k}_\mathrm{e}-\mathbf{P}/2 - \mathbf{K}+\mathbf{Q}} - \varphi^{\mathrm{1Sy}}_{\mathbf{k}_\mathrm{e}-\mathbf{P}/2} \right] \,\delta_{\mathbf{k}_\mathrm{e}+\mathbf{k}_\mathrm{h};\mathbf{K-Q+P}}\,.\nonumber
 \end{eqnarray}
Here, $\varphi_\mathbf{k}^i$ are the exciton envelope functions and $V_\mathrm{eff}(\mathbf{K})$ is the effective Coulomb interaction, both in momentum space. The $\langle u_{\mathrm{7v},\mathbf{q}}|(\mathbf{K}-\mathbf{Q})\cdot \mathbf{p}| u_{\mathrm{8c},\mathbf{q}}\rangle$ depicts the dipole transition matrix element between the $\Gamma_8^-$ conduction and the $\Gamma_7^+$ valence band.

Expressing the transition probability for the phonon-assisted (pa) process in Fermi's golden rule yields
\begin{eqnarray}\label{eq:fs05} 
&& \Gamma_\mathrm{Auger}^\mathrm{(pa)} = \frac{2\pi}{\hbar} \sum_{\mathbf{Q},\mathbf{K},\mathbf{P},\mathbf{k_e},\mathbf{k_h}} \left| M_\mathrm{1Sb}^\mathrm{(pa)}(\mathbf{Q},\mathbf{K},\mathbf{P},\mathbf{k_e},\mathbf{k_h}) \right|^2 \\
 &&\times\delta \left( E_\mathrm{1Sy}(\mathbf{K})\! +\! E_\mathrm{1Sy}(\mathbf{P})\! -\! E_\mathrm{6c}(\mathbf{k_e})\! -\! E_\mathrm{7v}(\mathbf{k_h})\! -\! \hbar\omega_\lambda (\mathbf{Q}) \right). \nonumber
\end{eqnarray}
At low temperatures, exciton momentum is considerably smaller than the momentum of the ionised particles due to the transferred energy being in the order of the gap energy, hence justifying the approximation $\mathbf{K},\mathbf{P}\rightarrow 0$. Then Eq.\ (\ref{eq:fs05}) simplifies to three sums, from which one is eliminated by the Kronecker delta of Eq.\ (\ref{eq:fs04}). The remaining sums are solved numerically. The transition matrix element as well as the momentum dependent deformation potential are known from fits of the $\Gamma_3^-$ phonon assisted absorption band edge \cite{schoene2017}. The resulting Auger coefficient is
\begin{align}\label{eq:fs06}
 A_\mathrm{Auger}^\mathrm{(pa)} \; = \; \Omega\,\Gamma_\mathrm{Auger}^\mathrm{(pa)} \; = \; 8.62\times 10^{-20} \,\mathrm{cm^3 ns^{-1}} \,,
\end{align}
which is closer to experimental values than the one calculated in \cite{baym1996}, but still several orders of magnitude below the expected value. However, it should be considered that this derivation is made for an unstressed crystal which, e.g., drastically inhibits the paraexciton recombination channel. For excitons in a trapped system the result of (\ref{eq:fs06}) should be regarded as a lower boundary for the possible Auger coefficient.

\subsubsection{Rate equations}
We assume that all electron-hole pairs rebind to form new excitons, which are randomly distributed among the four possible exciton states (one para- and three orthoexcitons). Therefore, one quarter of the newly formed excitons is fed into the paraexciton and half into the orthoexciton component, since the latter stands for ortho($+$)- and ortho($-$)excitons. In this setup the pump laser can not create ortho($0$)excitons. Therefore, the rebinding of electron-hole pairs is the only source of ortho($0$)excitons in the system. Due to the repulsive potential they are forced out of the trap. From the gradient of the potential one can estimate that it takes approximately $\unit{15}{\nano\second}$ for the ortho($0$)excitons to reach the fringe of the trap, if they start from its centre. The conversion lifetime on the other hand is about $\unit{4-5}{\nano\second}$. Hence, the ortho($0$)excitons are assumed to almost completely convert to paraexcitons on their way out of the trap. To take this into account we refeed one quarter of the newly created excitons into the paraexcitons smeared out over the whole trap with energy corresponding to the splitting.

The newly formed excitons are assumed to be at rest and are relaxed into partial local equilibrium first before being introduced into the hydrodynamic equations. Their initial energy corresponds to their binding energy of $E_\mathrm{B}^\mathrm{P}=\unit{151.36}{\milli\electronvolt}$ for para- and $E_\mathrm{B}^\mathrm{O}=\unit{139.24}{\milli\electronvolt}$ for orthoexcitons. The densities ($\tilde{n}_\mathrm{Auger}^i$) and energies ($E_\mathrm{Auger}^i$) after the initial relaxation are calculated in the same fashion as in the case of excitons created by the pump laser (by solving a homogeneous Boltzmann equation). Hence, the moments are given by 
\begin{eqnarray}
 \label{eq:AugerMoments}
 \Gamma^{(0,i)}_\mathrm{Auger} &=& -2 A_{ii}\tilde{n}_i^2(\mathbf{r},t) - A_{ij} \tilde{n}_i(\mathbf{r},t) \tilde{n}_j(\mathbf{r},t) \nonumber \\ &&+ \tilde{n}^i_\mathrm{Auger}(\mathbf{r},t) \, , \nonumber \\ 
 \mathbf{\Gamma}^{(1,i)}_\mathrm{Auger} &=& -m\tilde{n}_i(\mathbf{r},t)\mathbf{v}_i(\mathbf{r},t) [2A_{ii}\tilde{n}_i(\mathbf{r},t) - A_{ij} \tilde{n}_j(\mathbf{r},t)]  \, , \nonumber \\ 
 \Gamma^{(2,i)}_\mathrm{Auger} &=& -2A_{ii}\tilde{n}_i(\mathbf{r},t) E_i(\mathbf{r},t) -A_{ij}\tilde{n}_j(\mathbf{r},t) E_i(\mathbf{r},t) \nonumber \\ &&+ E_\mathrm{Auger}^i \, . 
\end{eqnarray}
For the Auger rates $A_{ij}$ we use the values reported in Ref.\ \cite{stolz2012} $A_\mathrm{PP}=\unit{2 \times 10^{-18}}{\centi\meter^3/\nano\second}$, $A_\mathrm{OO}=\unit{4.9 \times 10^{-17}}{\centi\meter^3/\nano\second}$ and $A_\mathrm{PO}=(A_\mathrm{PP}+A_\mathrm{OO})/2$.

When the two-body decay is attributed to the formation of biexcitons, Eqs.\ (\ref{eq:AugerMoments}) have to be modified. Instead of Auger rates, temperature-dependent capture coefficients $C_{ij}$ are used to model the process \cite{wolfe2014}:
\begin{eqnarray}
 C_\mathrm{PP}&=&\frac{\unit{1.4 \times 10^{-14}}{\centi\meter^3\kelvin/\nano\second}}{T_\mathrm{P}}\,, \nonumber \\
 C_\mathrm{OO}&=&\frac{\unit{4.7 \times 10^{-15}}{\centi\meter^3\kelvin/\nano\second}}{T_\mathrm{O}}\,, \quad
 C_\mathrm{PO}=0\,.
\end{eqnarray}
Excitons created by rebinding of electron-hole pairs are treated the same way as before. Therefore, the moments take the form 
\begin{eqnarray}
 \label{eq:BiExcMoments}
 \Gamma^{(0,i)}_\mathrm{Biexc} &=& -2 C_{ii}\tilde{n}_i^2(\mathbf{r},t) + \tilde{n}^i_\mathrm{Biexc}(\mathbf{r},t) \, , \nonumber \\ 
 \mathbf{\Gamma}^{(1,i)}_\mathrm{Biexc} &=& -2C_{ii}m\tilde{n}^2_i(\mathbf{r},t) \mathbf{v}_i(\mathbf{r},t)   \, , \nonumber \\ 
 \Gamma^{(2,i)}_\mathrm{Biexc} &=& -2C_{ii}\tilde{n}_i(\mathbf{r},t) E_i(\mathbf{r},t)+ E_\mathrm{Biexc}^i \, . 
\end{eqnarray}

\section{Results}
 
For the calculations we use cw excitation and $a^s_\mathrm{PP}=2.1 \, a_X$, $a^s_\mathrm{OO} =a^s_\mathrm{PO} =2/3a^s_\mathrm{PP}$ \cite{shumway01} with a Bohr radius of $a_X=\unit{0.7}{\nano\meter}$.

For the discussion it is useful to introduce the exciton temperature in the trap centre $T^0_i$ and the mean exciton temperature    
\begin{equation}
 \langle T_i \rangle = \frac{1}{N_i} \int d\mathbf{r} T_i(\mathbf{r},t) \tilde{n}_i(\mathbf{r},t) \, .
\end{equation}
Other important temperatures are that of the helium bath $T_\mathrm{B}$, that of the phonons (of the crystal lattice) $T_\mathrm{Ph}$ and that one extracted from fitting the experimental spectra $T_\mathrm{S}$.

In the following we first present and discuss some general results using the Auger effect as the two-body decay mechanism. Afterwards we compare results using Auger effect, biexciton formation and no two-body loss mechanism. Last, we compare the theoretical results of our model with experimental data.

\subsection{Stationary state}

\begin{figure}
 \vspace*{0.2cm}
 
 \includegraphics[width=\linewidth]{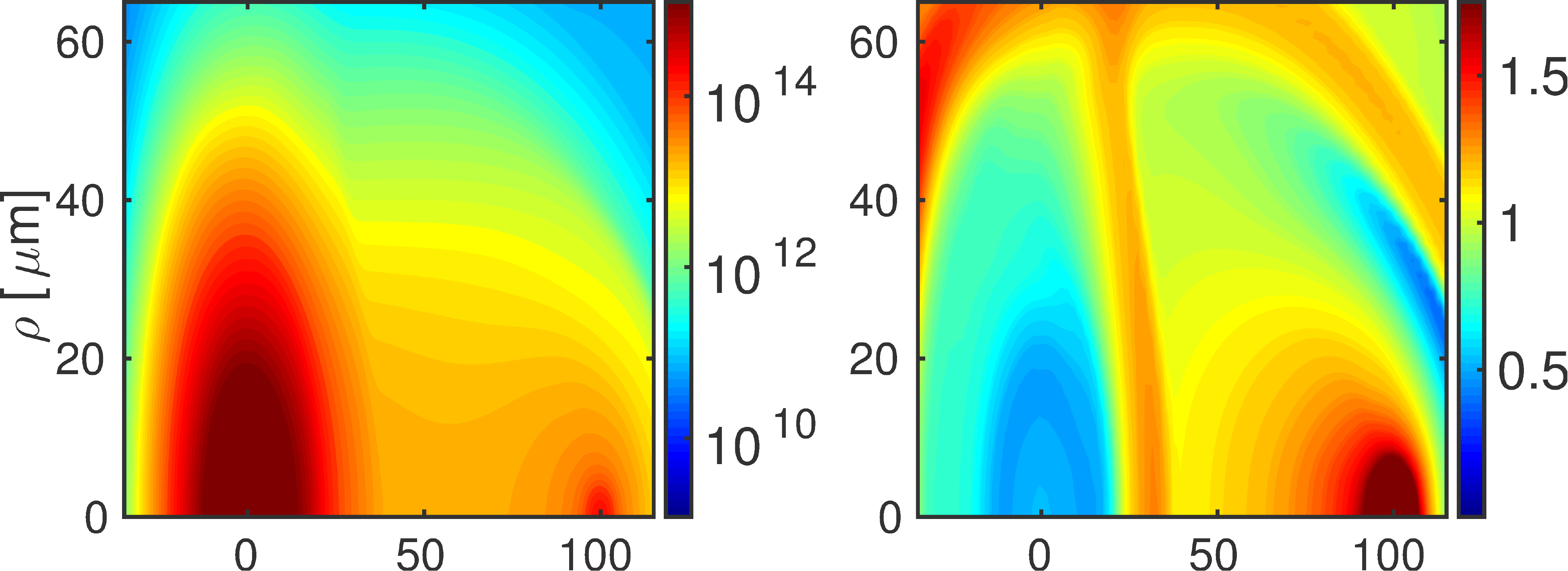}
 \vspace*{-0.25cm}
 
 \includegraphics[width=\linewidth]{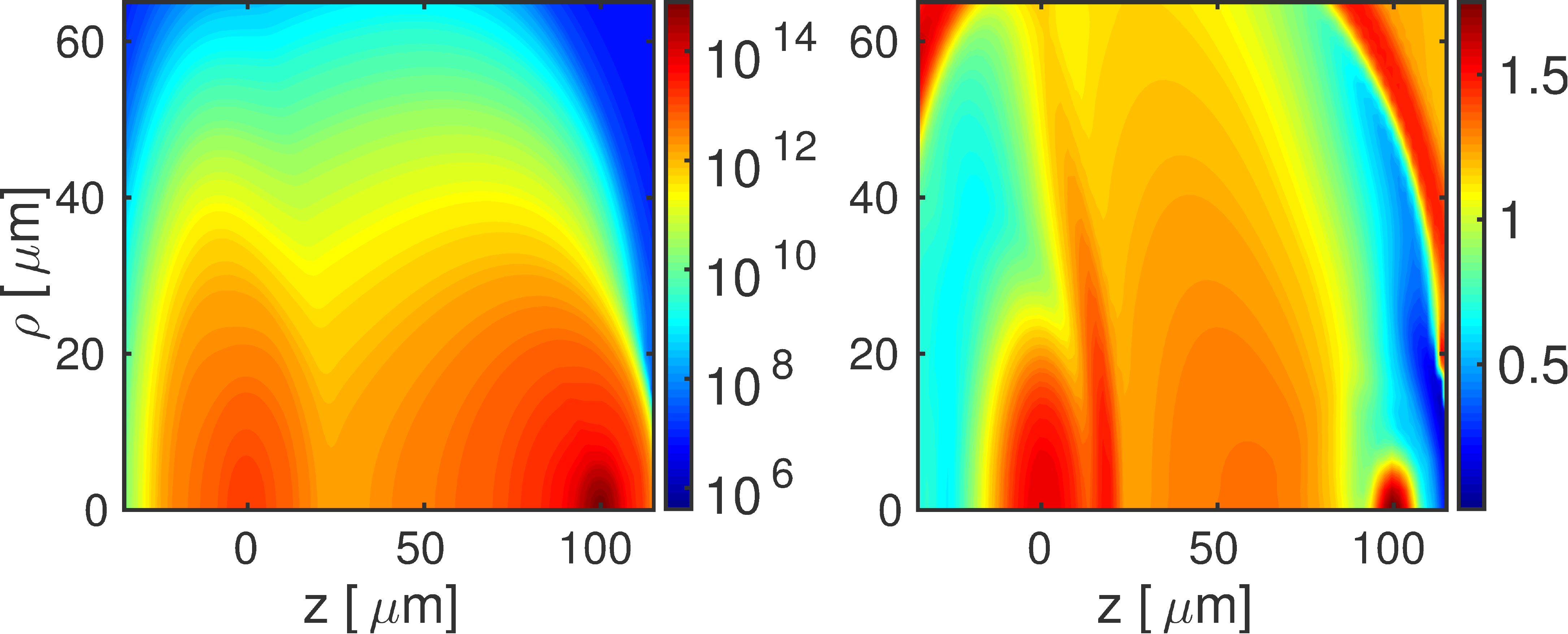}
 \caption{Density in $\centi\meter^{-3}$ (left column) and temperature in $\kelvin$ (right column) of the para- (top row) and orthoexcitons (bottom row) in the stationary state with $T_\mathrm{Ph}=\unit{0.25}{\kelvin}$ and $P_\mathrm{L}=\unit{69.52}{\micro\watt}$ as a function of $\rho$ (radial direction perpendicular to applied strain) and $z$ (direction of the strain).}
 \label{fig:DichteTemp2D}
\end{figure}

\begin{figure}
 \includegraphics[width=\linewidth]{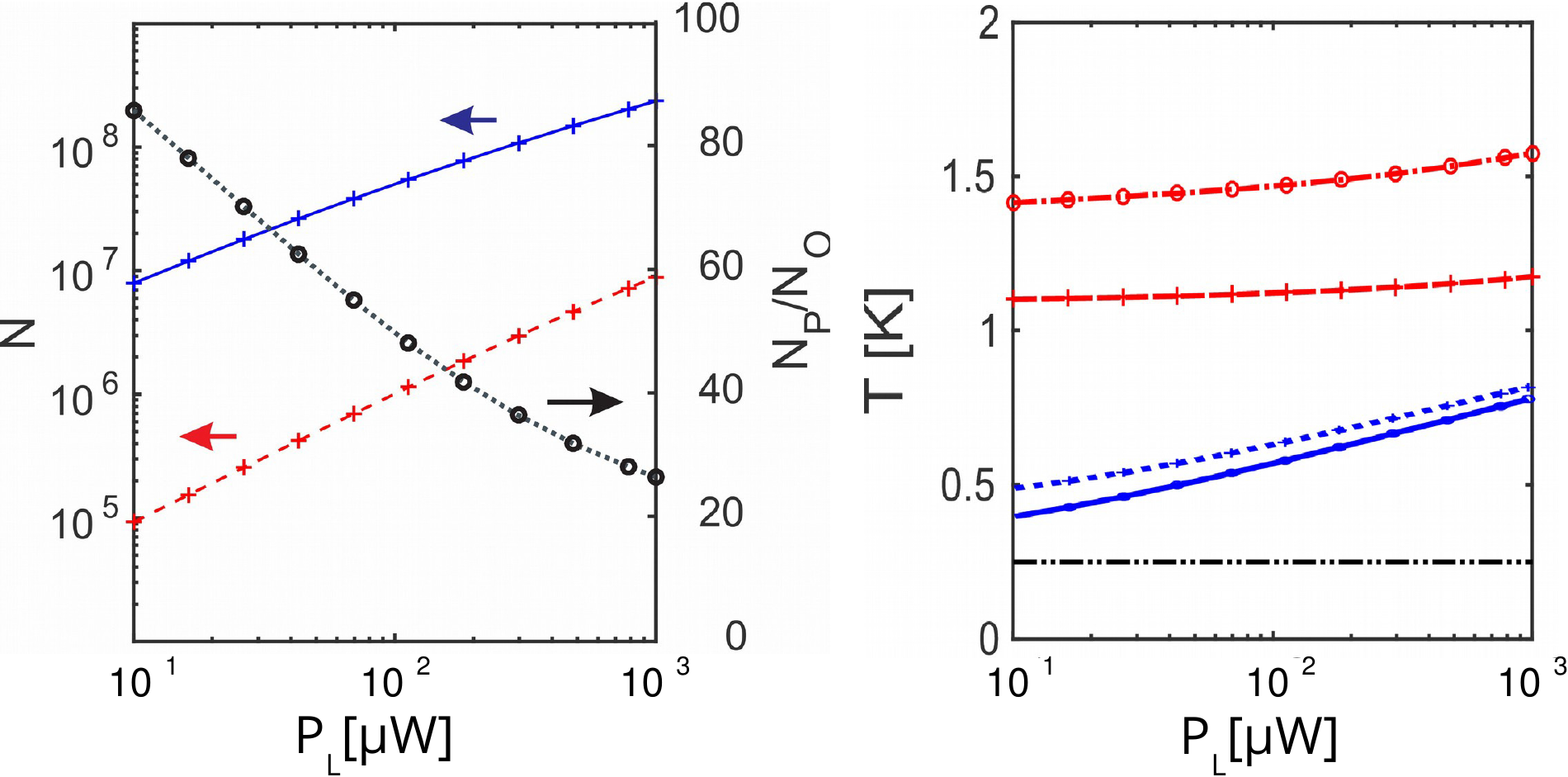}
%  \includegraphics[width=\linewidth]{NvonP.pdf}
%  \vspace*{-0.25cm}
%  
%  \includegraphics[width=\linewidth]{TvonP.pdf}
 \caption{Various quantities in the stationary state as function of the pump power $P_\mathrm{L}$ at $T_\mathrm{Ph}=\unit{0.25}{\kelvin}$ (black dashed line). Left: particle number of para- (blue solid) and orthoexcitons (red dashed), and ratio of para- to orthoexciton particle numbers (black dotted). Right: mean exciton temperature $\langle T_i \rangle$ (crosses) and exciton temperature in the trap centre $T^0_i$ (circles and ovals) for para- (blue, solid and short-dashed) and orthoexcitons (red, dash-dotted and long-dashed).}
 \label{fig:NTvonP}
\end{figure}

A typical result for temperature and density of the ortho- and paraexcitons is shown in Fig.\ \ref{fig:DichteTemp2D}. The two exciton species behave quite differently. The situation for the paraexcitons is straightforward. Hot excitons with temperatures of up to $\unit{3.5}{\kelvin}$ are created around $z=\unit{100}{\micro\meter}$ due to conversion of laser excited orthoexcitons. From there, they drift towards the trap centre, while gradually cooling down. They accumulate inside the trap and reach temperatures of around $\unit{0.5}{\kelvin}$. The highest density of the paraexcitons can be found in the trap centre. Since the effective lifetime of the orthoexcitons is much shorter, their situation is different. Most of the orthoexcitons created by the laser do not reach the trap centre. The highest density can be found in the centre of the laser spot. The orthoexcitons inside the trap mainly originate from recombining electron-hole pairs and, therefore, are very hot. The orthoexciton temperature inside the trap is around $\unit{1.5}{\kelvin}$, a factor of 3 higher than the paraexciton temperature.  
As shown in Fig.\ \ref{fig:NTvonP}, there are much more paraexcitons in the system than orthoexcitons for all pump powers considered here. However, the ratio of para- to orthoexcitons declines as the pump power increases. This is due to the growing influence of the Auger effect. The destroyed excitons are mainly paraexcitons but half of the rebinding electron-hole pairs become orthoexcitons, hence, the balance between ortho- and paraexcitons shifts in the system. The growing influence of the Auger effect is also visible in the development of the temperatures. For the paraexcitons the mean temperature as well as the temperature in the trap centre increase strongly, while their difference shrinks. This is due to the creation of high-energy excitons by the Auger effect, which acts as local heating where the density is high. The orthoexciton temperatures do not change drastically since they are already quite hot and the cooling by phonons is much more efficient at high temperatures.    

\subsection{Two-body loss mechanism}

In this section we analyse how our results depend on the implementation of the two-body loss mechanism. We consider the Auger effect and biexciton formation using the parameters given in Sec.\ \ref{sec:collision} and for reference the case of no two-body loss mechanism.   

\begin{figure}
 \includegraphics[width=\linewidth]{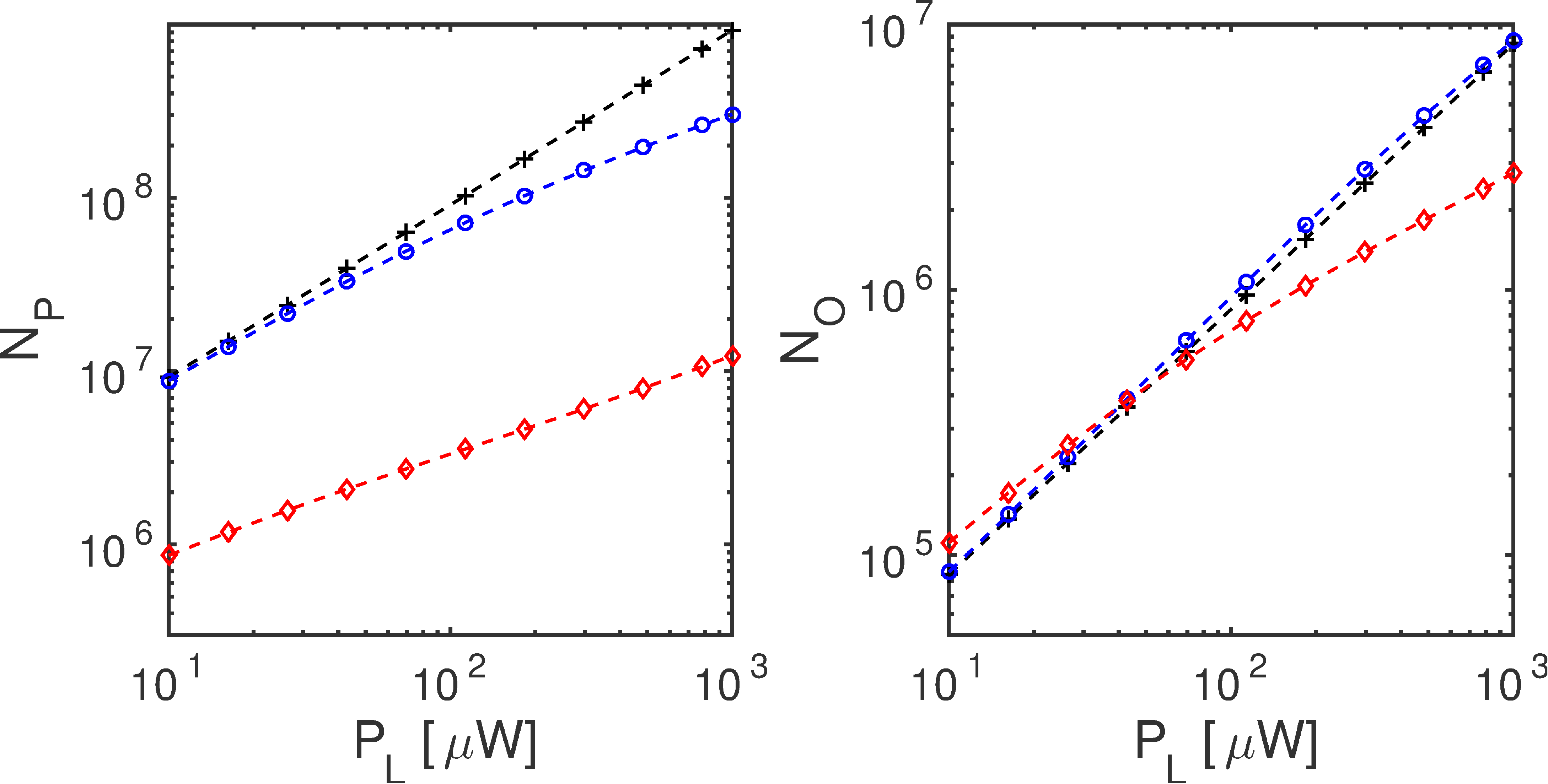}
 \vspace*{-0.25cm}
 
 \includegraphics[width=\linewidth]{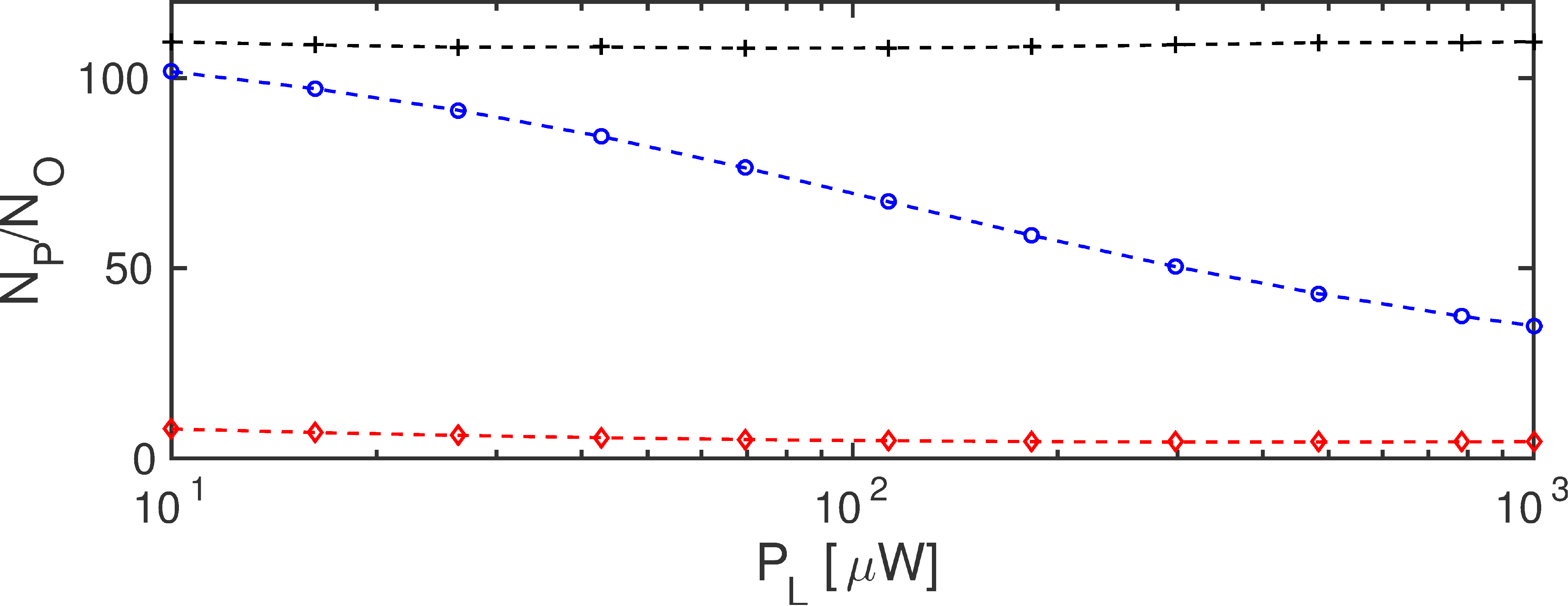}
 \caption{Particle number of para- (top left) and orthoexcitons (top right) and the ratio of para- to orthoexciton numbers in the stationary state as a function of pump power $P_\mathrm{L}$ using Auger decay (blue circles), biexciton formation (red diamonds) and without a two-body loss mechanism (black crosses) with $T_\mathrm{Ph}=\unit{1.0}{\kelvin}$.}
 \label{fig:NVervonP}
\end{figure}

\begin{figure}
 \includegraphics[width=\linewidth]{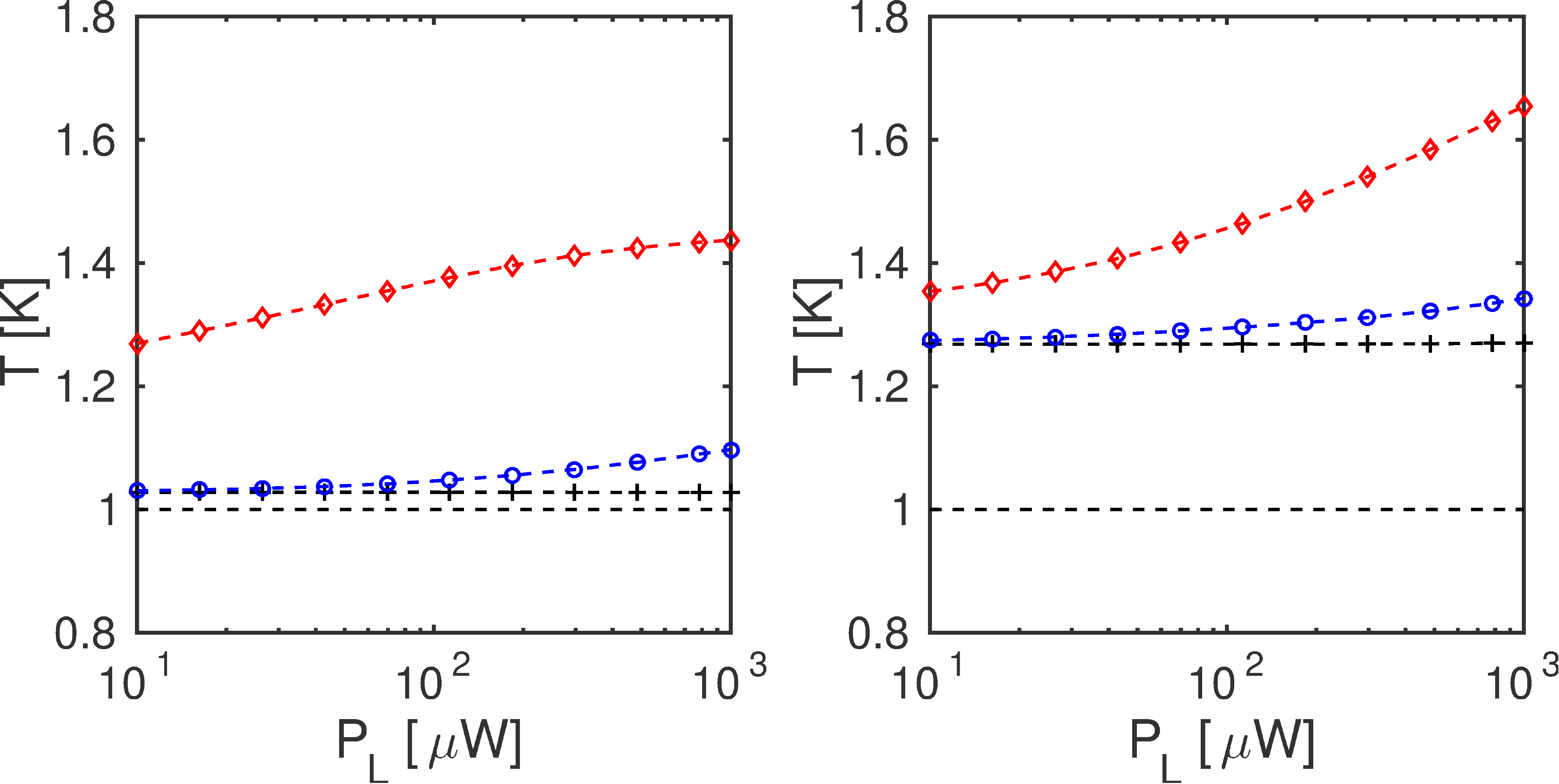}
 \caption{Mean exciton temperature $\langle T_i \rangle$ for para- (left) and orthoexcitons (right) for the same situations as in Fig.\ \ref{fig:NVervonP}.}
 \label{fig:TvervonP}
\end{figure}

\begin{figure}
 \includegraphics[width=0.9\linewidth]{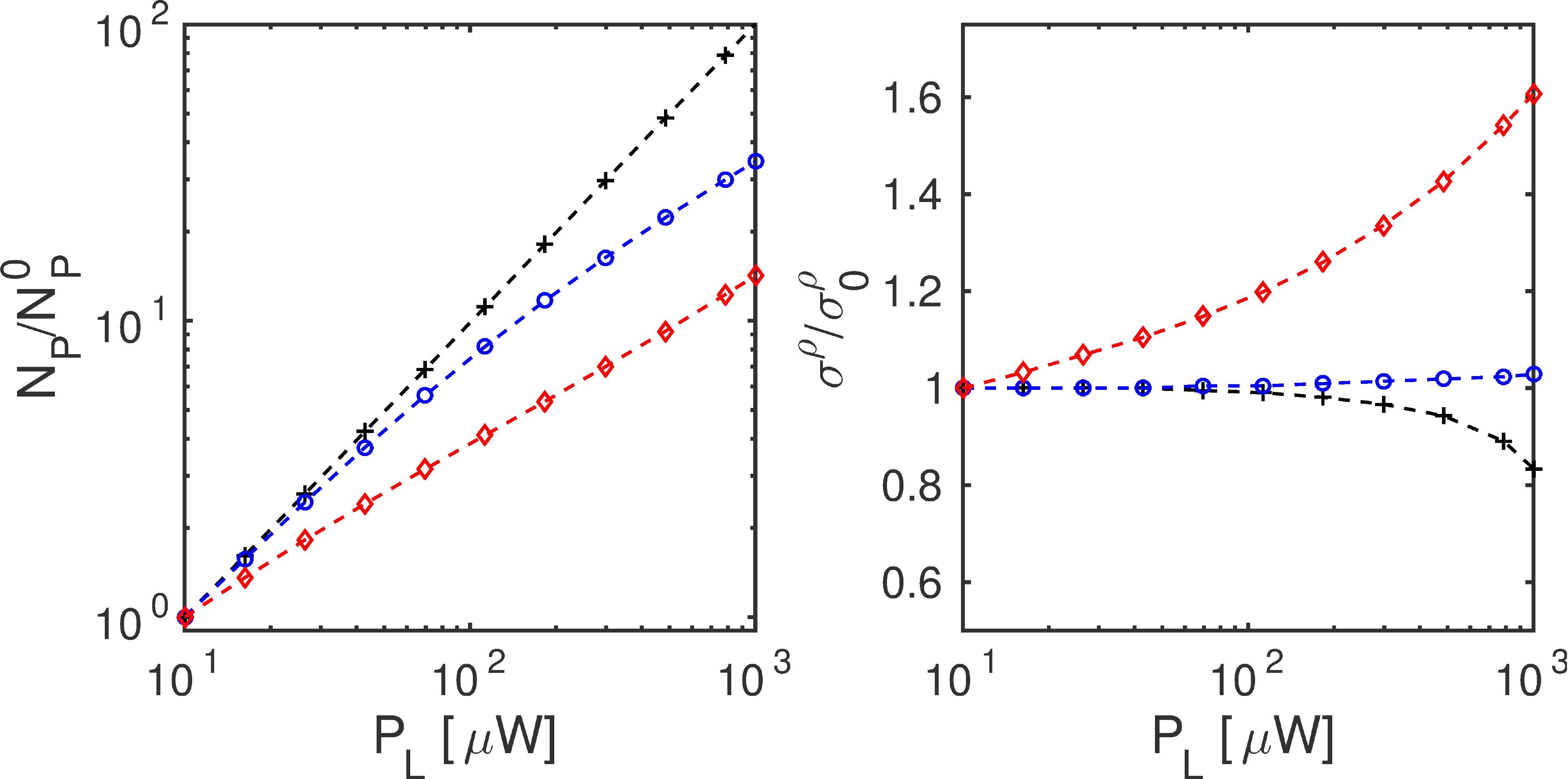}
 \caption{Normalised number of paraexcitons and extension of the thermal cloud in $\rho$-direction (full width at half maximum) for the same situations as in Fig.\ \ref{fig:NVervonP}.}
 \label{fig:VerNorm}
\end{figure}

The exciton numbers and the ratio of para- to orthoexcitons are given in Fig.\ \ref{fig:NVervonP}. Obviously, the loss due to the Auger effect is marginal over a wide range of pumping powers. Only at high laser powers the number of paraexcitons in the Auger effect model drops significantly. Using the biexciton model leads to paraexciton numbers one order of magnitude lower compared to the other cases, while the orthoexciton numbers only differ at higher pumping powers. The para to ortho fraction behaves differently for all cases. While the ratio stays almost constant using no two-body loss mechanism, it drops significantly using the Auger effect. For the biexciton model the ratio starts out much lower and drops slightly. This qualitatively different behaviour is also reflected in the mean temperature of the excitons depicted in Fig.\ \ref{fig:TvervonP}. The discrepancy of the bath temperature and the exciton temperature without two-body loss mechanism is due to the heat introduced into the system by the laser and the ortho-para conversion. The increase in temperature towards higher laser powers using the Auger effect is due to the hot excitons created by the latter. The temperatures in the biexciton model are over all pumping powers considerably higher. These strong differences between the two results should be observable in actual experiments. However, a comparison with experimental results is difficult since only the luminescence spectrum is available.
%Particle number and luminescence intensity are related in the simplest case by an unknown luminescence coefficient. In order to eliminate it,
Assuming a linear relation between particle number and luminescence intensity, one can normalise the intensities to the value of the lowest pumping power. Doing the same for the paraexciton numbers results in the left graph of Fig.\ \ref{fig:VerNorm}. Again the three cases behave quite differently. Another quantity that should be observable in actual experiments is the extension of the thermal cloud. We define it as the full width between the two points of half maximum along the $\rho$- or $z$-direction $\sigma^{\rho/z}$. The normalised result for the paraexcitons $\sigma^\rho/\sigma^\rho_0$ is shown in Fig.\ \ref{fig:VerNorm}. The extension of the thermal cloud also displays a completely different behaviour for the three cases. Hence, comparing the theoretical results of our model with experimental data should be helpful to determine how realistic the parameters are.     

\subsection{Comparison with experiments}

Comparing the experimentally determined luminescence intensities of the para- and orthoexcitons \cite{stolz2012}, one might be able to differentiate between these qualitatively different behaviours depicted in Fig.\ \ref{fig:NVervonP}.

\begin{figure}
 \includegraphics[width=\linewidth]{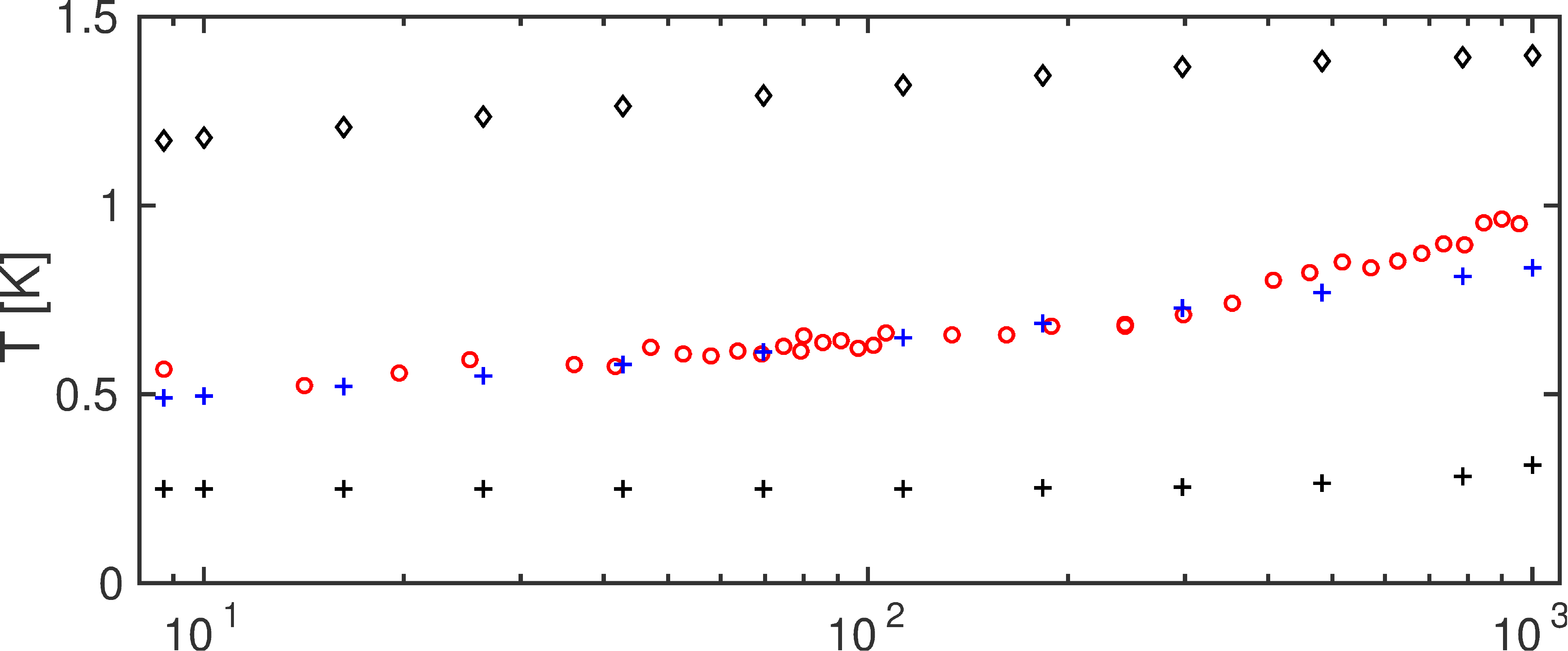}
 \includegraphics[width=\linewidth]{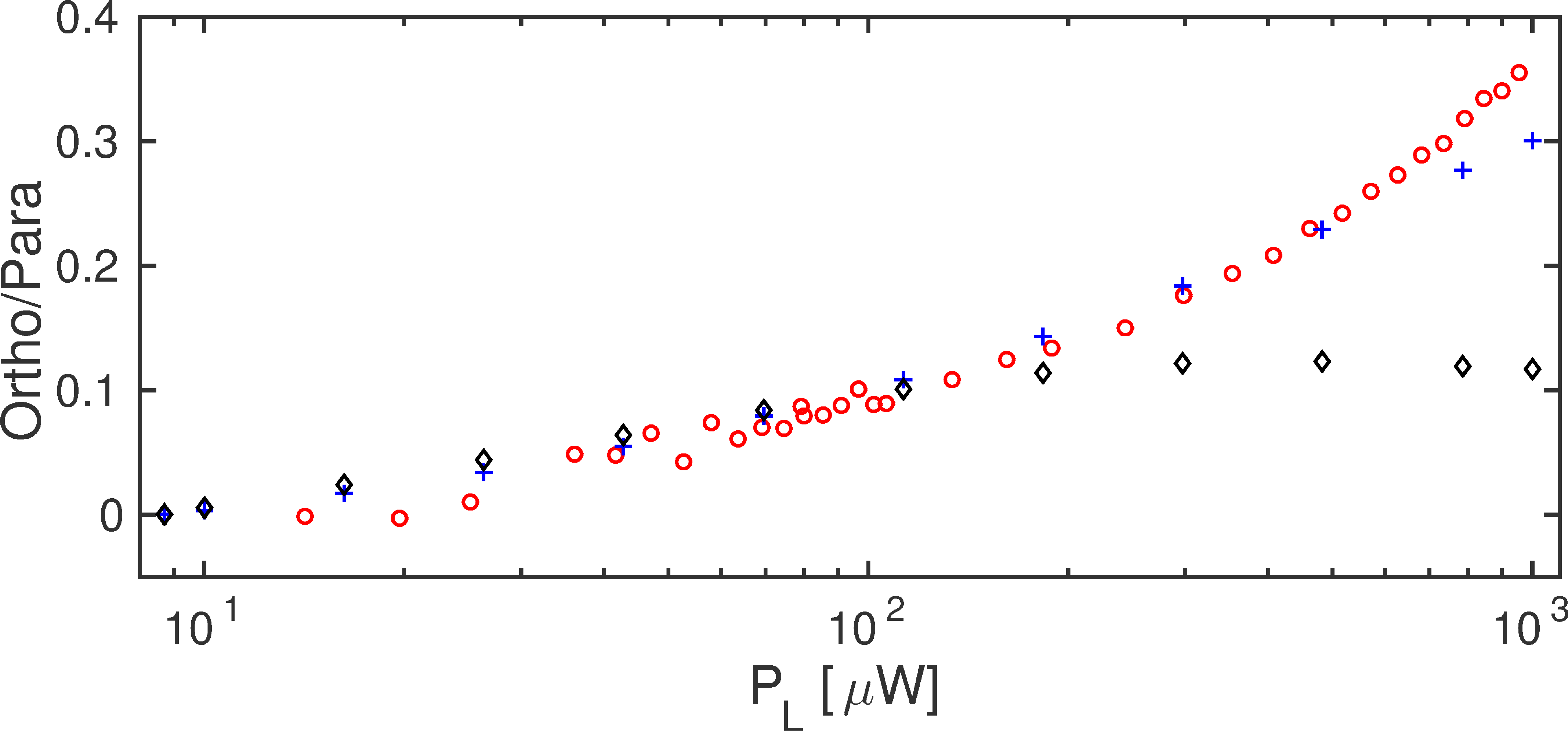}
 \caption{Top: mean paraexciton temperature $\langle T_P \rangle$ using biexciton formation (black diamonds) and Auger effect (blue crosses) compared to the spectral temperature determined from experimental data. The black crosses represent the helium bath temperature. Bottom: ratio of ortho- to paraexciton numbers (symbols as above) compared to the ratio of the integrated luminescence intensity of the ortho- and paraexcitons (experimental data).}
 \label{fig:VergExp}
\end{figure}

The trap depth for the paraexcitons in the experiments was $\unit{3.5}{\milli\electronvolt}$. The helium bath temperature was measured and ranged between 0.26 and $\unit{0.31}{\kelvin}$ depending on the pumping power. For our calculations we assumed the crystal to be in equilibrium with the bath $T_\mathrm{Ph}=T_\mathrm{B}$. In the top part of Fig.\ \ref{fig:VergExp} we compare the mean temperature of the paraexcitons according to theory with the spectral temperature obtained from experimental data. The temperatures predicted using the biexciton model are considerably higher than the spectral temperatures. On the other hand, the results using the Auger effect agree quite well with the experimental data. However, a growing discrepancy appears for higher pump powers. This could be due to a heating of the crystal that might be stronger than the rise in the bath temperature suggests. 

A comparison of the ratio of para- and orthoexcitons between experiment and theory is difficult for two reasons. First, the luminescence intensity of the orthoexcitons is indistinguishable from the noise for the lowest pumping powers. Second, there is an unknown factor between the intensities of the different species. To account for that we subtract the first value from the theoretical data and multiply the result with a factor to adapt the theoretical to the experimental data. Note that this is the only fit parameter. The results in the lower part of Fig.\ \ref{fig:VergExp} clearly show a strong discrepancy between the experimental data and the results using the biexciton formation model. The latter show a qualitatively different behaviour for high pumping powers than the experimental results. The results using the Auger effect provide the correct qualitative behaviour even though there are some deviations at low and high pumping powers.

\begin{figure}
 \includegraphics[width=\linewidth]{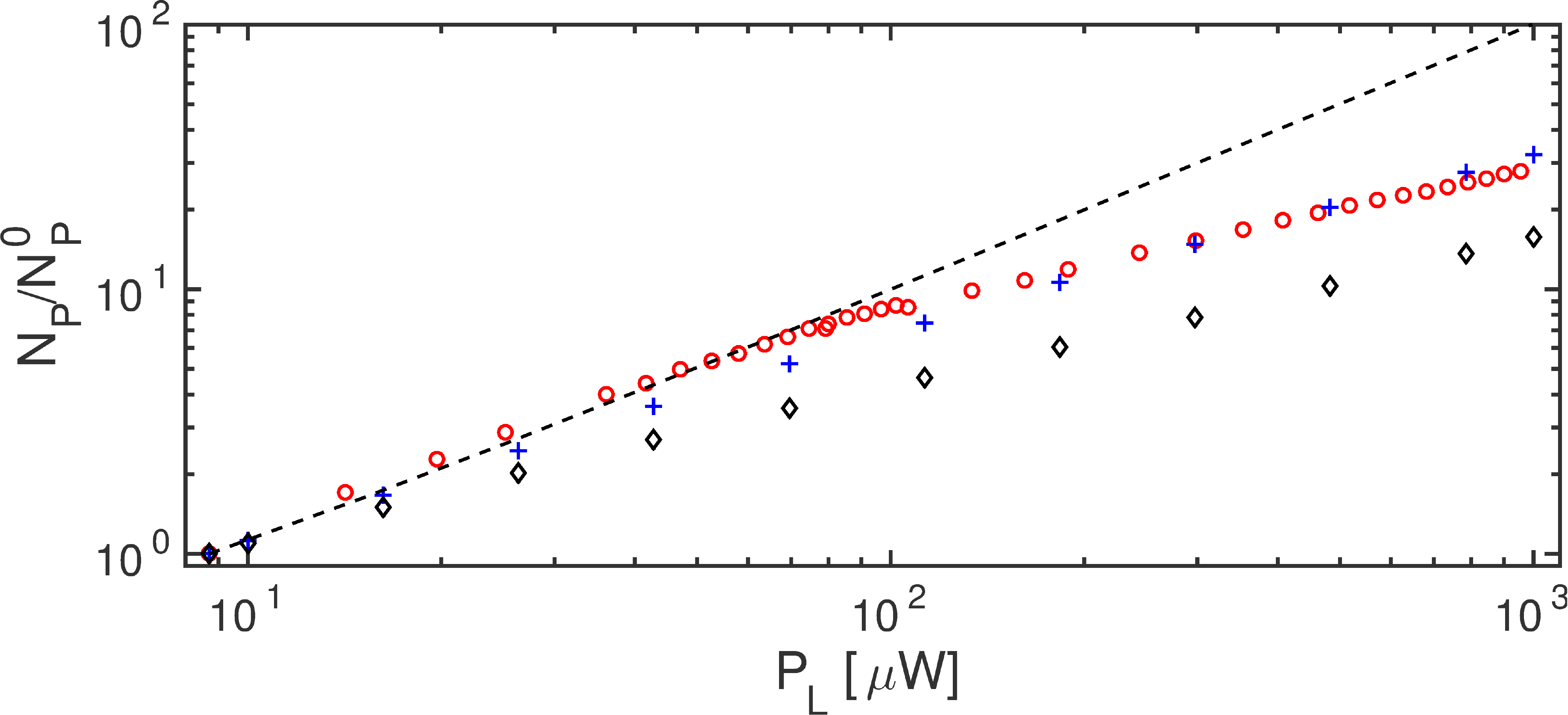}
 \caption{Normalised paraexciton numbers using biexciton formation (black diamonds) and Auger effect (blue crosses) compared to the normalised integrated luminescence intensity (red circles; experimental data). The black dashed line represents the expected behaviour without two-body loss mechanism.}
 \label{fig:VergExp1}
\end{figure}

In Fig.\ \ref{fig:VergExp1} we compare the normalised paraexciton number with the normalised integrated luminescence intensity from the experimental data. In the first half of the plot the experimental data follow the dashed line quite well, indicating a very weak or no influence of the two-body loss mechanism. For these pumping powers both theoretical models behave quite differently from the experimental data with the results of the Auger effect being closer to the experimental data. However, in the second half the experimental values agree quite well with the Auger effect model. The results for the biexciton model are slightly lower compared to the other two. In conclusion, the results using the Auger effect display a reasonable agreement with the experimental data, while the biexciton model seems to be quantitatively and in one case even qualitatively off. However, we want to emphasise that it cannot be concluded from this result that the proposed mechanism is wrong. It simply means that the suggested capture coefficients seem to be too high to explain the experimental results at these low temperatures.  

\subsection{Auger effect and BEC}

\begin{figure}
 \includegraphics[width=\linewidth]{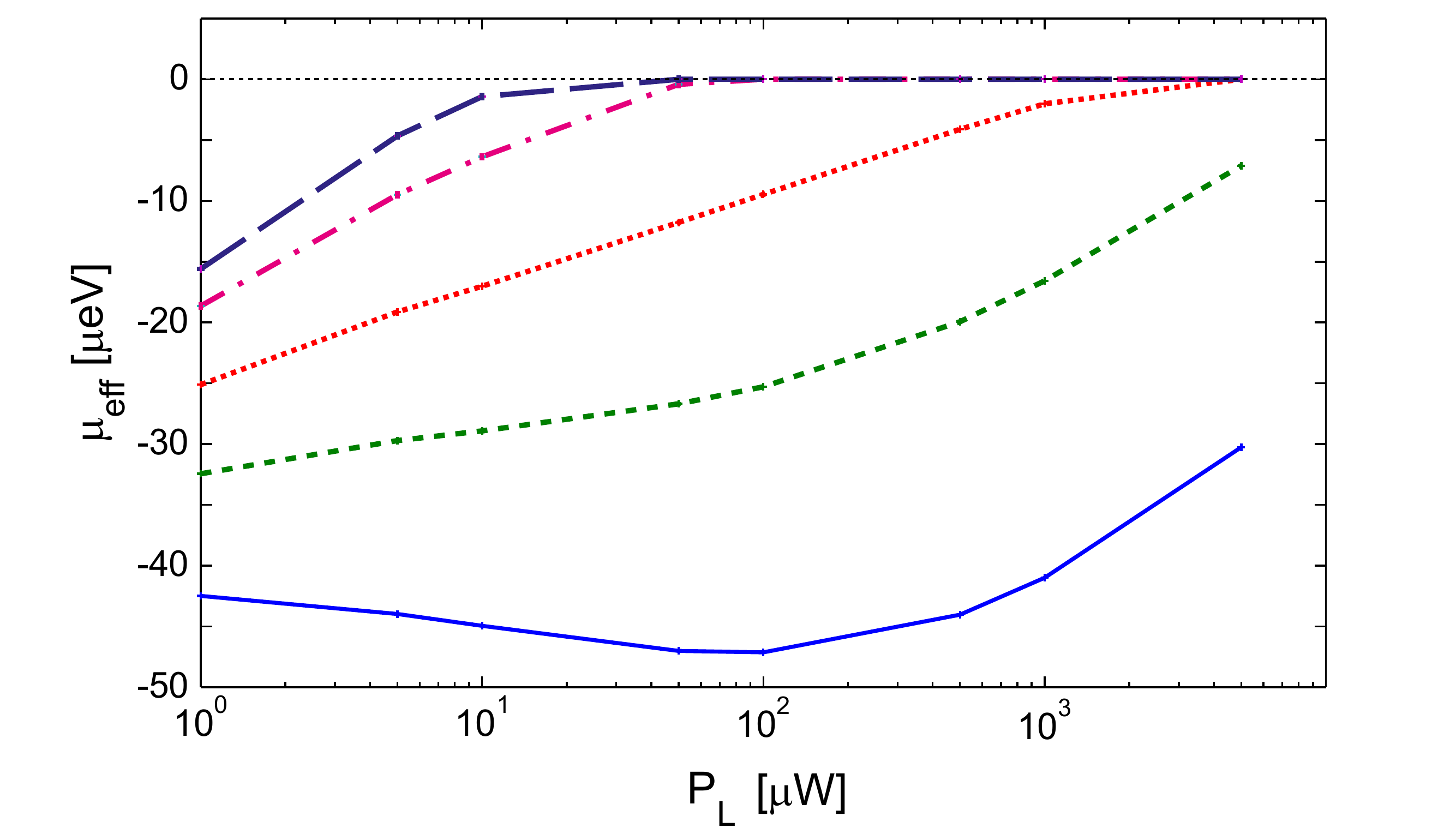}
 \caption{Effective chemical potential $\mu_\mathrm{eff}$ vs.\ laser power $P_\mathrm{L}$ for various Auger coefficients $A$. $A_0=\unit{2.0\times10^{-18}}{\centi\meter^3/\nano\second}$ and $A/A_0=1$ (blue solid line), 0.5 (green dashed), 0.25 (red dotted), 0.1 (magenta dash-dotted), and 0.05 (dark blue long-dashed). The bath temperature is $T_\mathrm{B}=\unit{0.037}{\kelvin}$.}
 \label{fig:myeff}
\end{figure}

Although modeling the two-body loss mechanism using the Auger effect is able to reproduce experimental results qualitatively quite well, a quantitative description of the Auger effect in a strained crystal is still pending. In particular, the possibility of reaching the conditions for BEC, i.e., high enough densities and low enough temperatures, strongly depends on the absolute strength of the Auger coefficient. This is illustrated by Fig.\ \ref{fig:myeff} \cite{sobkowiak2014a}. The vanishing of the effective chemical potential of the paraexcitons $\mu_\mathrm{eff}\equiv\mu-U=0$ marks the condensation boundary. There is obviously a critical value of the Auger coefficient of about $5\times 10^{-19}$ cm$^3$/ns above which no BEC is possible within reasonable laser powers.

\section{Conclusion and Outlook}

% -presented model to describe the experiments using an effective two component system \newline
% -using cylindrical symmetry we can directly input experimental data for trap and so on \newline
% -discussed possible quantities for comparison of experiment and theory without further assumptions \newline
% -compared experimental and theoretical data \newline
% -Auger effect yields reasonable agreement however to simple \newline
% -biexciton model to high capture coefficients \newline
% -spectrum still missing \newline

The present theoretical approach is capable to describe the dynamics of a trapped multicomponent exciton gas in Cu$_2$O.
% The system is described to be effectively two-component. The inclusion of the orthoexcitons extends our previous analysis described in Refs.\ \cite{sobkowiak2014,sobkowiak2015}.
For an effective two-component system we analysed the different behaviour of para- and orthoexcitons in the course of the drift towards the trap centre. We showed in particular that the number of paraexcitons always exceeds that of the orthoexcitons substantially while the temperature of the latter species is by a factor of 3 higher than that of the former one. Particle number as well as temperature show the growing influence of the two-body loss mechanism with increasing pump power.

The comparison of two models for the two-body decay process -- Auger effect vs.\ transient biexciton formation -- shows reasonable agreement of theoretical and experimental data for the first model while there are significant deviations for the latter one. The biexciton model, however, cannot be completely ruled out by this analysis, but only the given high capture coefficients.

Since theoretical and experimental results are compared on the basis of the light emission by the excitons, a comprehensive theory of the excitonic decay luminescence is still in demand.

\begin{acknowledgments}
We would like to thank G.\ Manzke, W.-D.\ Kraeft, and Th.\ Bornath for many fruitful discussions. This work was supported by the Deutsche Forschungsgemeinschaft via Collaborative Research Center SFB 652 (project B14).
\end{acknowledgments}

\end{document}